\documentclass[aps,prl,twocolumn,showpacs,superscriptaddress,groupedaddress]{revtex4} 

\usepackage{amsmath}
\usepackage{amssymb}
\usepackage{graphicx}
\usepackage{bm}

\begin{document}

\title{Multiple Scattering and Plasmon Resonance in the Intermediate Regime}
\author{Bo Liu}
\affiliation{Department of Physics, Harvard University, Cambridge, Massachusetts 02138, USA}
\author{Eric J. Heller}
\affiliation{Department of Physics, Harvard University, Cambridge, Massachusetts 02138, USA} 
\affiliation{Department of Chemistry and Chemical Biology, Harvard University, Cambridge, Massachusetts 02138, USA} 
\date{\today}

\begin{abstract}
The collective excitation of the conduction electrons in subwavelength structures gives rise to the Localized Surface Plasmon(LSP). The system consisting of two such LSPs, known as the dimer system,is of fundamental interest and is being actively investigated in the literature. Three regimes have been previously identified and they are the photonic regime, the strong coupling regime and the quantum tunneling regime. In this Letter, we propose a new regime for this intriguing systems, the intermediate regime. In this new regime, the quasistatic approximation, which is widely used to study such LSP systems, fails to capture the main physics: the multiple scattering of the electromagnetic waves between the two LSPs, which significantly modifies the properties of the resonant modes in the system. This intermediate regime provides a new route to explore in plasmonics, where controlling both the excited plasmon modes and the damping rates are of paramount significance.
\end{abstract}


\maketitle
The collective excitation of conduction electrons in subwavelength structures is known as the Localized Surface Plasmon(LSP)\cite{a1}. Such plasmon modes have been intensively studied using noble metal nanoparticles\cite{a2,a3,a4,a5}. More recently, the possibility of building terahertz metamaterials supporting such LSP modes is explored using graphene microribbons\cite{a6} and microdisks\cite{a7}. LSP holds promise for applications in ultrasensitive biosensing\cite{a8}, nano-optical tweezers\cite{a9} and improved photovoltaic devices\cite{a10}. 

\begin{figure}
\begin{center}
\includegraphics[width=7cm]{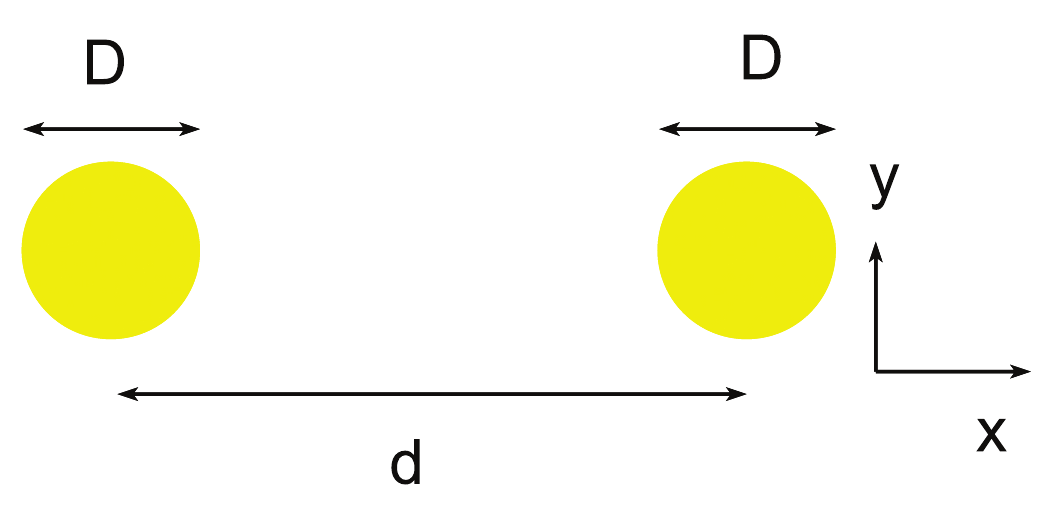}
\end{center}
\caption{\label{setup}\footnotesize{(color online) Schematic of the system setup. Two nanoparticles with diameter D are separated by a distance d. }}
\end{figure}

\begin{figure*}
\begin{center}
\includegraphics[width=18cm]{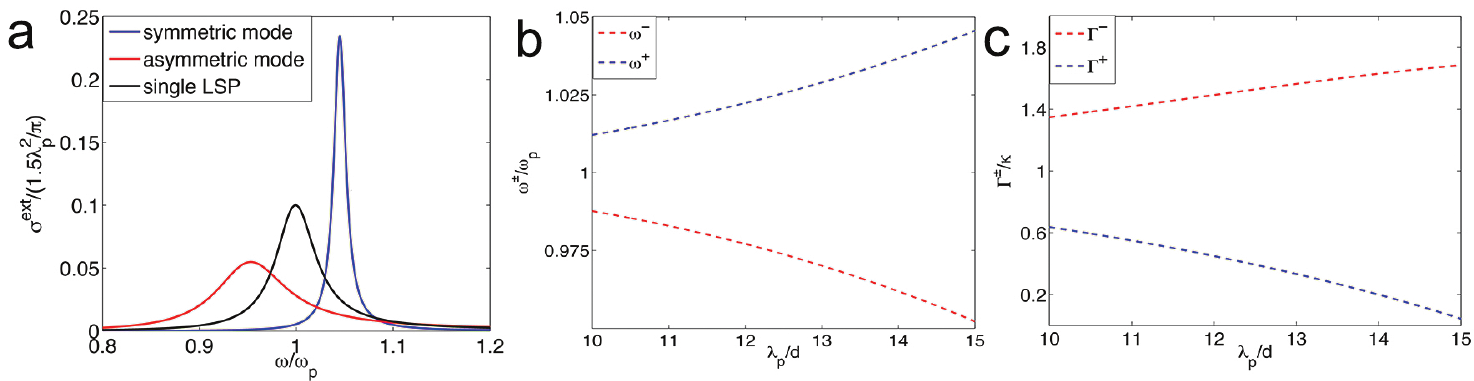}
\end{center}
\caption{(color online)\label{damping}\footnotesize{ ({\bf a})Extinction cross sections for the single LSP(black), the antisymmetric mode(red), and the symmetric mode(blue). In this plot,  $\lambda_p/d$ is chosen to be 15 and the cross section for the symmetric mode is multiplied by a factor of 0.1. ({\bf b}) The resonance frequencies of the antisymmetric mode and the symmetric mode as a function of $\lambda_p/d$. The red dashed line corresponds to  the antisymmetric mode while the blue dashed line corresponds to the symmetric mode. The resonance frequencies are measured in the unit of the LSP resonance frequency $\omega_p$. ({\bf c})The damping rates of the antisymmetric mode and the symmetric mode as a function of $\lambda_p/d$. The red dashed corresponds to the antisymmetric mode while the blue dashed line corresponds to the symmetric mode. The damping rates are measured in terms of the damping rate of the single LSP damping rate $\kappa$. In all three plots, $\kappa_r/\kappa$ is fixed to be $10\%$ and $\omega_p/\kappa$ is assumed to be 20.}} 
\end{figure*}

LSP arises when a subwavelength metal nanosphere is illuminated by light. This problem can be solved analytically within the quasistatic approximation\cite{a1}, where one ignores the phase retardation and solves the problem using electrostatics. Due to its simplicity, the quasistatic approximation has been widely applied to study systems involving LSPs \cite{a3,a4,a5,a6,b4} and powerful theoretical methods including transformation optics\cite{b2,b3} and the hybridization model\cite{a5,b5} are developed under this approximation. However, by ignoring the phase retardation, one misses many potentially interesting effects arising from multiple scattering. In this Letter, we study such an effect in systems consisting of two weakly interacting LSPs, which is popularly known as the dimer system. 

The dimer system is being actively explored in the literature due to its richness of fundamental physics and many potential applications . Three regimes have been identified:

\begin{enumerate}
  \item  The Photonic regime\cite{d1}, where d $\sim \lambda$.  d, as defined in Fig.\ref{setup}, is the center to center distance between the two nanoparticles supporting the LSPs  and $\lambda$ is the wavelength of the incident light.
  \item The strong coupling regime\cite{b3}, where $D<d<2D$ and D is the diameter of the nanoparticles.
  \item The quantum tunneling regime\cite{b10}, where $d-D$ $\sim$ 0.1nm(atomic spacing).
\end{enumerate}

In this Letter, we propose a new regime for this dimer system, the intermediate regime, which arises when $2D<d<0.1\lambda$.  This regime was previously named the weak coupling regime in Ref.\cite{b3} and it was argued within the quasistatic approximation that the dimer system in this regime should exhibit the same behavior as individual LSPs, meaning that only a single symmetric mode can be observed. This quasistatic argument would be correct if the two nanoparticles are non-resonant scatters of the incident light. However, the LSPs are by nature resonant scattering modes\cite{a1} with scattering cross section scaling as $\lambda^2$\cite{a1, a11}. For nanoparticles with D<20nm, this scattering cross section can be 10000 times larger than their physical sizes\cite{a2}. If two such nanoparticles are placed within one wavelength, one can expect multiple scattering to yield new interesting physics in certain regimes. In quantum scattering theory, multiple scattering between two resonant scatters gives rise to the proximity resonance\cite{a11}. Due to the distinctive resonant properties of LSPs, multiple scattering displays a different signature in this plasmonic dimer system.

In the following, we first consider the dimer system illuminated by an incoming plane wave polarized in the y direction and propagating in the x direction. For nanoparticles with $D<20nm$, the LSPs can be well described by two resonant dipoles, $\vec{p}_1(\vec{r}_1)$, $\vec{p}_2(\vec{r}_2)$\cite{a1,b3,d2}, where $\vec{r}_1,\vec{r}_2$ are the positions of the two nanoparticles.  These two dipoles have to satisfy the following self-consistent equations:
\begin{equation} 
\begin{aligned}
\vec{p}_1(\vec{r}_1)&=\alpha(\omega)[\vec{E}_0(\vec{r}_1)+G(\vec{r}_1-\vec{r}_2)\vec{p}_2(\vec{r}_2)]\\
\vec{p}_2(\vec{r}_2)&=\alpha(\omega)[\vec{E}_0(\vec{r}_2)+G(\vec{r}_2-\vec{r}_1)\vec{p}_1(\vec{r}_1)]
\label{incom}
\end{aligned}
\end{equation},
where $\alpha(\omega)$ is the dynamic electric polarizability tensor, $\vec{E}_0(\vec{r})$ is the incoming wave and $G(\vec{r})$ is the interaction tensor defined as\cite{d2,a12}
\begin{equation} 
\begin{aligned}
G(\vec{r})=(k^2+\bigtriangledown \bigtriangledown)\frac{e^{ikr}}{r},
\end{aligned}
\end{equation}
and k is the light momentum in free space.

A direct expansion of (\ref{incom}) yields

\begin{equation} 
\begin{aligned}
&\vec{p}_1(\vec{r}_1)=\alpha(\omega)\vec{E}_0(\vec{r}_1)+\alpha(\omega)G(\vec{r}_1-\vec{r}_2)\alpha(\omega)\vec{E}_0(\vec{r}_2)\\
&+\alpha(\omega)G(\vec{r}_1-\vec{r}_2)\alpha(\omega)G(\vec{r}_2-\vec{r}_1)\alpha(\omega)\vec{E}_0(\vec{r}_1)+\cdots
\end{aligned}
\end{equation}
, from which it is clear that $\vec{p}_1(\vec{r}_1)$ includes contributions from both the incoming wave and all the waves that are scattered by the nanoparticle at $\vec{r}_2$ and eventually return to $\vec{r}_1$. These include all the possible multiple scattering paths between the two nanoparticles.

For our setup, the solutions to (\ref{incom}) can be written as the sum of a symmetric part and an antisymmetric part:
\begin{equation} 
\begin{aligned}
&\vec{p}_s=\frac{1}{2}\frac{\alpha_{yy}(\omega)[\vec{E}_0(\vec{r}_1)+\vec{E}_0(\vec{r}_2)]}{1-\alpha_{yy}(\omega)G_{yy}(\vec{r}_1-\vec{r}_2)}\\ 
&\vec{p}_{as}=\frac{1}{2}\frac{\alpha_{yy}(\omega)[\vec{E}_0(\vec{r}_1)-\vec{E}_0(\vec{r}_2)]}{1+\alpha_{yy}(\omega)G_{yy}(\vec{r}_2-\vec{r}_1)}
\label{sas}
\end{aligned}
\end{equation},
where $\alpha_{yy}(\omega)$ and $G_{yy}(\vec{r})$ are the diagonal elements corresponding to the y directions, of the polarizability tensor and the interaction tensor respectively. 

With this definition, the solutions are given by
\begin{equation} 
\begin{aligned}
\vec{p}_1(\vec{r}_1)=\vec{p}_s+\vec{p}_{as},\ \ \vec{p}_2(\vec{r}_2)=\vec{p}_s-\vec{p}_{as}.
\end{aligned}
\end{equation}

 As is clear from (\ref{sas}), new resonant modes arise when $Re[\alpha_{yy}(\omega)G_{yy}(\vec{r}_2-\vec{r}_1)]=\pm 1$, corresponding to a symmetric and an antisymmetric mode respectively. 
The physical origin of these two resonant modes are different from those in the photonic regime and the strong coupling regime. In the photonic regime, an antisymmetric mode could arise if $d=\lambda/2$, which corresponds to a phase matching condition. However, the antisymmetric mode in this intermediate regime arises when $d<0.1\lambda$. In this case, the phase accumulation due to multiple scattering enables the excitation of the antisymmetric mode for subwavelength separation between the two nanoparticles. In the strong coupling regime, the antisymmetric mode arises as a result of the hybridization of the individual dipole modes\cite{a5}, which requires the separation d to be smaller than twice of the diameter of the nanoparticles\cite{b3}. This condition corresponds to $\lambda/d>27$ for $D=15nm$\cite{a2} and this mode gains strength as this ratio increases\cite{b10}. However, the antisymmetric mode in the intermediate regime can arise for $\lambda/d<=15$ and loses strength as this ratio increases. More importantly, the antisymmetric mode in the strong coupling regime can arise within the quasistatic approximation\cite{b3} where phase retardation is completely ignored. However, as we will demonstrate below, the antisymmetric mode in the intermediate regime will completely disappear in the quasistatic limit, which clearly points to a different origin from that in the strong coupling regime.

\begin{figure}
\begin{center}
\includegraphics[width=8cm]{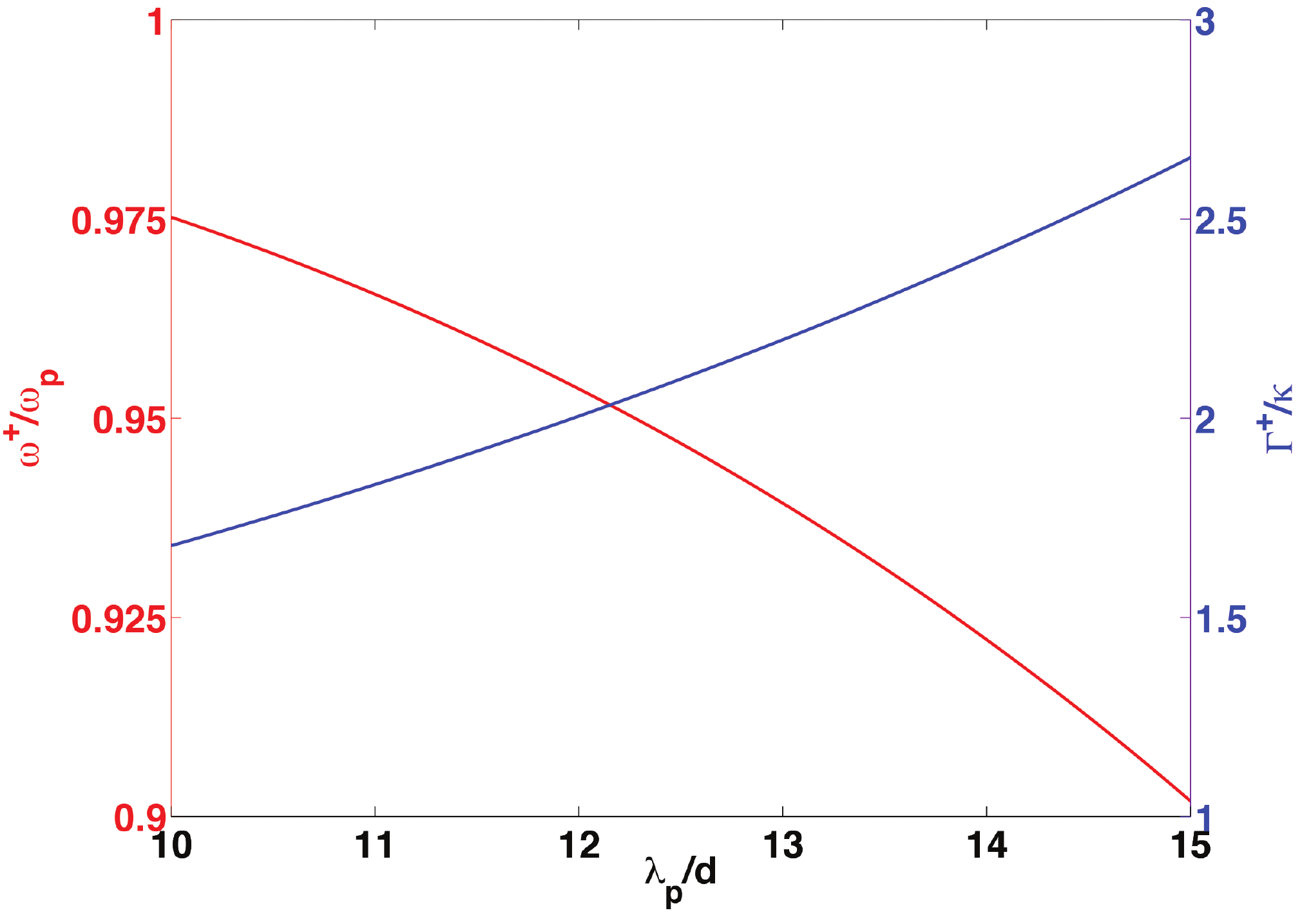}
\end{center}
\caption{(color online)\label{damping2}\footnotesize{Left: The resonance frequency of the symmetric mode as a function of $\lambda_p/d$, measured in the unit of $\omega_p$. Right: The damping rate of the symmetric mode as a function of $\lambda_p/d$,  measured in the unit of $\kappa$.  }}
\end{figure}

For nanoparticles supporting LSPs, $\alpha_{yy}(\omega)$ can be written as
\begin{equation} 
\begin{aligned}
\alpha_{yy}(\omega)=\frac{3c^3\kappa_r}{2\omega_p^2}\frac{1}{\omega_p^2-\omega^2-i\kappa\omega^3/\omega_p^2},
\label{polar}
\end{aligned}
\end{equation}
where $\omega_p$ is the single LSP resonance frequency, $\kappa$ is the total damping rate and $\kappa_r$ is the radiative contribution to $\kappa$. This form has the merits of satisfying both the optical theorem and causality in the absence of absorption\cite{a13,a14}.

If we define $\omega_o=\sqrt{\frac{3c^3\kappa_r}{2\omega_p^2d^3}}$ and keep only the leading terms, we find the following simplified expressions for the resonant frequency and damping rate for the antisymmetric mode:
\begin{equation} 
\begin{aligned}
\omega^{-}&=\sqrt{\omega_p^2-\omega_o^2\cos\frac{\omega_pd}{c}}\\
\Gamma^{-}
&=\kappa+\frac{\omega_o^2}{\omega^{-}}\sin\frac{\omega_pd}{c}-\frac{ \kappa \omega_{o}^2}{\omega_p^2}\cos\frac{\omega_pd}{c}. 
\end{aligned}
\end{equation}

For the symmetric mode, the results are
\begin{equation} 
\begin{aligned}
\omega^{+}&=\sqrt{\omega_p^2+\omega_o^2\cos\frac{\omega_pd}{c}}\\
\Gamma^{+}
&=\kappa+\frac{\kappa \omega_o^2}{\omega_p^2}\cos\frac{\omega_pd}{c}-\frac{\omega_o^2}{\omega^{+}}\sin\frac{\omega_pd}{c}.
\end{aligned}
\end{equation}

The above formula implies that, under certain conditions, the damping rate of the antisymmetric mode can increase above the single LSP damping rate $\kappa$, while the damping rate of the symmetric mode can drop below $\kappa$. These conditions are 
$\frac{\omega_p d}{c}>\frac{\omega^{-}}{\omega_p^2} \kappa$, $\frac{\omega_p d}{c}>\frac{\omega^{+}}{\omega_p^2}\kappa$,
which can be easily satisfied by gold nanoparticles as shown below.

In Fig.\ref{damping}, we plot the extinction cross section for each mode\cite{d3}, the resonance frequencies $\omega^{\pm}/\omega_p$ and the damping rates $\Gamma_{\pm}/\kappa$ as a function of $\lambda_p/d$, where $\lambda_p=\omega_p/c$ is the wavelength of the exciting light at the single LSP frequency. In these calculations, $\kappa_r/\kappa$ is fixed to be $10\%$ and $\omega_p/\kappa$ is assumed to be 20, which are chosen based on a previous experiment on gold nanoparticles with $D=15nm$\cite{a2}. When $\lambda_p/d=15$, the damping rate of the antisymmetric mode is boosted by a factor of 1.8, while the damping rate of the symmetric mode is reduced by a factor of 5.  Combined with a 10$\%\ \omega_p$ splitting in the resonant frequencies, the two modes can be easily distinguished in an experimental setup. Even though the antisymmetric mode is weaker than the symmetric mode, it is still greatly enhanced by multiple scattering. As one can see from Fig.\ref{damping}a, the resonance peak for the antisymmetric mode is only smaller than the single LSP resonance peak by a factor of about two. Since the single LSP resonance is strong enough to enable single molecule detection\cite{d4}, the antisymmetric mode in this intermediate regime is clearly strong enough to have observable effects.

As mentioned above, what distinguishes this antisymmetric mode in the intermediate regime from that in the strong coupling regime is its reliance on the phase retardation in the incident field. If the phase retardation is removed by changing the polarization of the incident light($\vec{E}_0//x$ and $\vec{k}//y$), the antisymmetric mode is completely suppressed and only the symmetric mode can be excited. The resonance frequency and the damping rate for the symmetric mode in this case are found to be
\begin{equation} 
\begin{aligned}
\omega'^{+}&=\sqrt{\omega_p^2-2\omega_o^2\cos\frac{\omega_pd}{c}}\\
\Gamma'^{+}
&=\kappa-\frac{2\kappa \omega_o^2}{\omega_p^2}\cos\frac{\omega_pd}{c}+\frac{2\omega_o^2}{\omega'^{+}}\sin\frac{\omega_pd}{c}. 
\end{aligned}
\end{equation}

These relations are plotted in Figure \ref{damping2} using the same parameters as before. In this case, the symmetric mode instead displays enhanced damping.

To summarize, we introduce a new intermediate regime for studying the plasmon dimer system, where multiple scattering has the dominant effect. We show that one can excite the antisymmetric mode by direct light illumination. It was previously believed that this antisymmetric mode,important for plasmon-induced transparency\cite{b7} and three-dimensional plasmon rulers\cite{b8}, can't be excited by the incident light in this "weak coupling" regime and one has to resort to either electron beam\cite{c1} or designed structures with broken symmetry\cite{b1,b6, b7,b8} to activate it. More importantly, we show that multiple scattering can lead to significant changes in the damping rates of the plasmon modes in this new regime.  The damping rate of the symmetric mode, a highly radiating mode, can be reduced by a factor of five, while the damping rate of the antisymmetric mode, a subradiant mode, is enhanced above the single LSP damping rate. This could be important for many practical applications, such as Surface-Enhanced Raman Scattering\cite{b9}, improved photovoltaic devices\cite{a10}, optical emitters\cite{b10,b11} and plasmon waveguides\cite{b12}. But most importantly, manipulating plasmon modes and their damping rate are of key importance in the field of plasmonics, and our results provide a new route to explore to this end.

We thank Prof. Federico Capasso for valuable discussions and acknowledge financial support from the U.S. Department of Energy under Grant DE-FG02-08ER46513.


\begin{thebibliography}{99}
\bibitem{d3}
The total extinction cross section for this system is found to be $\sigma^{tot}(\omega)=\sigma^{s}(\omega)+\sigma^{as}(\omega)$, where $\sigma^{s}(\omega)=\frac{8\pi \omega Im[\alpha_{yy}(\omega)]}{c\left|1-\alpha_{yy}(\omega)G_{yy}(\vec{r}_1-\vec{r}_2) \right |^2}$ and $\sigma^{as}(\omega)=\frac{8\pi \omega Im[\alpha_{yy}(\omega)]}{c\left|1+\alpha_{yy}(\omega)G_{yy}(\vec{r}_1-\vec{r}_2) \right |^2}$. We can define $\sigma^{s}(\omega)$ as the extinction cross section for the symmetric mode and $\sigma^{as}(\omega)$ as the extinction cross section for the antisymmetric mode.
\end{thebibliography}
\end{document}